\documentstyle[12pt]{article}
\def\beq{\begin{equation}}
\def\eeq{\end{equation}}

\def\bt{\beta}

\def\de{\delta}
\def\De{\Delta}

\def\te{\theta}

\def\lam{\lambda}

\def\ep{\epsilon}

\def\sq{\sqrt}

\def\l{\left (}
\def\r{\right )}
\def\fr{\frac} 
\def\la{\label}
\def\hs{\hspace}

\def\ran{\rangle}
\def\lan{\langle}
\def\tm{\times}
\def\tl{\tilde}

\begin{document}

\begin{titlepage}

\begin{center}
{\Large\bf   Bi-maximal Neutrino Mixing \\
And Anomalous Flavor ${\cal U}(1)$ 
}
\end{center}
\vspace{0.5cm}
\begin{center}
{\large Qaisar Shafi$^{a}$\footnote {E-mail address:
shafi@bartol.udel.edu} {}~and
{}~Zurab Tavartkiladze$^{b}$\footnote {E-mail address:
z\_tavart@osgf.ge} }
\vspace{0.5cm}

$^a${\em Bartol Research Institute, University of Delaware,
Newark, DE 19716, USA \\

$^b$ Institute of Physics, Georgian Academy of Sciences,
380077 Tbilisi, Georgia}\\
\end{center}

\vspace{1.0cm}

\begin{abstract}
 We describe how bi-maximal neutrino mixing can be realized 
in realistic models based on MSSM and SUSY  GUTs such as  $SU(5)$ and
$SO(10)$.  A crucial role is played by an anomalous ${\cal U}(1)$ flavor
symmetry, which also helps understand the observed charged fermion mass 
hierarchies and the magnitudes of the CKM matrix elements.  While in 
MSSM a variety of solutions for the solar neutrino puzzle are possible, 
SUSY $SU(5)$ and $SO(10)$ only permit the large mixing angle MSW solution.
Models in which the ${\cal U}(1)$ symmetry also mediates SUSY breaking 
allow both the large mixing angle and the 
low MSW solutions.

We also present renormalization group studies for the neutrino mass
matrix, generated through the ${\cal U}(1)$
flavor symmetry. Our analysis shows that renormalization  does not 
change the desirable picture of  bi-maximal mixing.  

\end{abstract}

\end{titlepage}

Despite its remarkable experimental successes there is little doubt 
that the standard model based on $SU(3)\times SU(2)\times U(1)$ must be
part of a more complete theory.  
Support for this statement comes from a variety of sources.  
Two that are particularly relevant for us here are the atmospheric 
(and solar) neutrino puzzles, as well as the well known flavor problem.  
We will attempt to resolve them within a unified framework such as 
provided by the minimal supersymmetric standard model (MSSM), 
$SU(5)$ or $SO(10)$, supplemented by a flavor ${\cal U}(1)$ symmetry.

In the charged fermion sector there are noticeable hierarchies within 
the charged
fermion Yukawa couplings and the CKM matrix elements. Since the mass 
of the top quark is close to the electroweak symmetry breaking scale
($\sim 100$~GeV), its Yukawa coupling is of order unity ($\lam_t \sim 1$).
As far as the Yukawa couplings of the $b$ quark and $\tau $ lepton are
concerned, their values could vary in a range 
$\lam_b \sim \lam_{\tau }\sim 10^{-2}-1 $, depending on the value of 
the MSSM parameter $\tan \bt $ ($\sim 1 - 60$). Introducing the
dimensionless parameter
$\ep \simeq 0.2$ (close to the Wolfenstein parameter $\lam $), 
one can express the observed hierarchies between the charged fermion
Yukawa couplings as follows:

\beq
\lambda_t\sim 1~,~~
\lambda_u :\lambda_c :\lambda_t \sim
\epsilon^6:\epsilon^3 :1~,
\la{ulam}
\eeq

\beq
\lam_b\sim \lam_{\tau}\sim \lam_t\fr{m_b}{m_t}\tan \bt ~,~~~
\lambda_d :\lambda_s :\lambda_b \sim
\epsilon^5:\epsilon^2 :1~,
\label{dlam}
\eeq

\beq
\lambda_e :\lambda_{\mu } :\lambda_{\tau } \sim
\epsilon^5:\epsilon^2 :1~,
\label{elam}
\eeq
while for the CKM matrix elements:

\beq
V_{us}\sim \epsilon~,~~~V_{cb}\sim \epsilon^2~,~~~
V_{ub}\sim \epsilon^3~.
\label{ckm}
\eeq
In constructing models, one should arrange for a natural understanding of
the hierarchies in (\ref{ulam})-(\ref{ckm}).

The latest atmospheric and solar neutrino data (see \cite{atm} and \cite{sol}
respectively) seem to provide convincing confidence in the phenomena of neutrino
oscillations. Ignoring the LSND data \cite{lsnd}, the atmospheric and solar
anomalies can be explained within the three states of active neutrinos.
In this paper we will study oscillation scenarios
without the sterile neutrinos, which in any case maybe disfavored by the data
\cite{sol, atm, akhm}. 

The atmospheric neutrino data suggest oscillations of $\nu_{\mu }$ into $\nu_{\tau }$,
with the following oscillation parameters:

$$
{\cal A}(\nu_{\mu }\to \nu_\tau )\equiv \sin^2 2\te_{\mu \tau}
\simeq 1~,
$$
\beq
\De m^2_{\rm atm}\simeq 3\cdot 10^{-3}~{\rm eV}^2~.
\la{atmdat}
\eeq

The solar neutrino anomaly seems consistent with a variety of oscillation scenarios, amongst which the
most likely seems to be large angle MSW (LAMSW) oscillation of $\nu_e$
into $\nu_{\mu , \tau }$ \cite{sol}, with the oscillation parameters:

$$
{\cal A}(\nu_e\to \nu_{\mu, \tau } )\equiv \sin^2 2\te_{e \mu , \tau}
\simeq 0.8~,
$$
\beq
\De m^2_{\hs{0.5mm}\rm sol}\simeq 2\cdot 10^{-5}~{\rm eV}^2~.
\la{LAMSWdat}
\eeq
The scenario of low MSW (LOW MSW) oscillations of solar neutrinos
require:

$$  
\sin^2 2\te_{e \mu , \tau}
\simeq 1.0~,
$$  
\beq
\De m^2_{\hs{0.5mm}\rm sol}\simeq 8\cdot 10^{-8}~{\rm eV}^2~,
\la{LOWMSWdat}
\eeq
while the small angle MSW (SA MSW) oscillations are realized with:

$$
\sin^2 2\te_{e \mu , \tau }
\simeq 5\cdot 10^{-3}~,
$$
\beq
\De m^2_{\hs{0.5mm}\rm sol}\simeq 5\cdot 10^{-6}~{\rm eV}^2~.
\la{smalMSW}
\eeq
Finally, the large angle vacuum oscillation (LAVO) solution
requires

$$
\sin^2 2\te_{e \mu , \tau }
\simeq 0.7~,
$$
\beq
\De m^2_{\hs{0.5mm}\rm sol}\simeq 8\cdot 10^{-11}~{\rm eV}^2~.
\la{LAVOdat}
\eeq

It is worth noting that within MSSM, the neutrinos acquire masses only through
non-renormalizable $d=5$ Planck scale operators $l_il_jh_u^2/M_P$ which,
for
$\lan h_u^0\ran \sim 100$~GeV, $M_P=2.4\cdot 10^{18}$~GeV (reduced Planck
mass)
give $m_{\nu_i }\sim 10^{-5}$~eV. Therefore, already for atmospheric
data (\ref{atmdat}) [and also for solar neutrino anomalies with 
(\ref{LAMSWdat})-(\ref{smalMSW})] we need physics beyond the MSSM.
In order to generate the appropriate neutrino masses, we will introduce
heavy right handed neutrino states ${\cal N}_i$. Then, the `light' left
handed neutrinos will acquire masses through the see-saw mechanism
\cite{seesaw}.

In building neutrino oscillation scenarios, the main challenge is to generate
desirable magnitudes for neutrino masses and their mixings. And to
understand why in some cases, the mixing angles are large (and even
maximal), while the quark CKM matrix elements (\ref{ckm}) are suppressed.
Below we will present a mechanism which successfully resolves all
of these problems.

Before proceeding to the model, let us study a specific
neutrino mass matrix texture, which provides bi-maximal neutrino 
mixing. Consider the mass matrix for three active neutrino flavors:

\beq
\begin{array}{ccc}
&  {\begin{array}{ccc}
\hspace{-5mm}~~ & \,\, & \,\,

\end{array}}\\ \vspace{2mm}
\begin{array}{c}
\\  \\
\end{array}\!\!\!\!\!\hat{M}_{\nu }=&{\left(\begin{array}{ccc}
\,\,0~~ &\,\,m_1~~ &
\,\,m_2
\\
\,\,m_1~~   &\,\,0~~  &
\,\,0
 \\
\,\,m_2~~ &\,\,0~~ &\,\,0
\end{array}\right) }~
\end{array}  \!\!~~~~~,
\label{nu}
\eeq
and assume that

\beq
m_1\sim m_2~
\la{scales}
\eeq
are real mass parameters.
Performing the transformation $U_1^{T}M_{\nu }U_1\equiv M_{\nu }'$, where 

\beq
\begin{array}{ccc}
U_1=~~
\!\!\!\!\!\!\!\!&{\left(\begin{array}{ccc}
\,\,1 &\,\,~~0 &
\,\,~~0
\\
\,\,0  &\,\,~~c_{\theta }
&\,\,~-s_{\theta }
\\
\,\, ~0 &\,\,~~
s_{\theta }  &\,\,~~c_{\theta }
\end{array}\right)}
\end{array}~,
\label{u1}
\eeq
\beq
s_{\theta }\equiv \sin \theta~,~~~c_{\theta }\equiv \cos \theta~,~~~
\tan \theta =\frac{m_2}{m_1}~,
\label{angles}
\eeq
the mass matrix acquires the off-diagonal form:

\begin{equation}
\begin{array}{ccc}
&  {\begin{array}{ccc}
\hspace{-5mm}~~ & \,\, & \,\,

\end{array}}\\ \vspace{2mm}
\begin{array}{c}
 \\  \\
 \end{array}\!\!\!\!\!\hat{M'}_{\nu }= &{\left(\begin{array}{ccc}
\,\,0~~ &\,\,m~~ &
\,\,0
\\
\,\,m~~   &\,\,0~~  &
\,\,0
\\
\,\,0 &\,\,0~~ &\,\,0
\end{array}\right) }~,
\end{array}  \!\!  ~~~~~~~m=\sqrt{m_1^2+m_2^2}~,
\label{nu2}
\eeq
which is diagonalized through transformations with maximal rotating
angles:

\beq
U_2^T\hat{M'}_{\nu }U_2\equiv \hat{M}_{\nu }^{\rm diag}=
{\rm Diag }(m~,~-m~,~0)~,
\label{tran3}
\eeq

\beq
\begin{array}{ccc}
U_2=~~
\!\!\!\!\!\!\!\!&{\left(\begin{array}{ccc}
\,\,\frac{1}{\sqrt{2}} &\,\,~-\frac{1}{\sqrt{2}} &
\,\,~~0
\\
\,\,\frac{1}{\sqrt{2}} &\,\,~~~~\frac{1}{\sqrt{2}}
&\,\,~~0
\\ 
\,\, ~0 &\,\,~~~~0  &\,\,~~1
\end{array}\right) }
\end{array}~.
\label{u2}
\eeq
The rotation matrix which connects the mass eigenstate basis to the flavor
one is given by:

\beq
\begin{array}{ccc}
U_{\nu }=U_1U_2=~~
\!\!\!\!\!\!\!\!&{\left(\begin{array}{ccc}
\,\,\frac{1}{\sqrt{2}} &\,\,-\frac{1}{\sqrt{2}} &
\,\,~~0
\\
\,\,\frac{1}{\sqrt{2}}c_{\theta }  &\,\,~~~~\frac{1}{\sqrt{2}}c_{\theta }
&\,\,-s_{\theta }
\\
\,\, ~~\frac{1}{\sqrt{2}}s_{\theta } &\,\,~~~~
\frac{1}{\sqrt{2}}s_{\theta }  &\,\,~~c_{\theta }
\end{array}\right) }
\end{array}~,
\label{lepckm}
\eeq
which is precisely a bi-maximal mixing matrix. Through (\ref{lepckm}), using the
expression for the oscillation amplitude

\beq
{\cal A}(\nu_{\alpha }\to \nu_{\beta })=
4\Sigma_{j<i}U_{\nu }^{\alpha j}U_{\nu }^{\alpha i}
U_{\nu }^{\beta j}U_{\nu }^{\beta i}~,
\label{defamp}
\eeq
($\alpha, \beta $ denote flavor indices and $i, j$
the mass eigenstates),
the oscillation amplitudes for the atmospheric 
and solar neutrinos are respectively

\beq
{\cal A}(\nu_{\mu }\to \nu_{\tau })\equiv \sin^2 2\te_{\mu \tau }=
\frac{4m_1^2m_2^2}{(m_1^2+m_2^2)^2}\sim 1~,
\la{ampatm}
\eeq

\beq
{\cal A}(\nu_e\to \nu_{\mu, \tau })\equiv \sin^2 2\te_{e \mu, \tau}=1~.
\label{ampsol}
\eeq
In (\ref{ampatm}) we have taken into account (\ref{scales}), 
while the second equation in (\ref{ampsol}) holds
without any assumptions. For $m_1\simeq m_2$ we will have 
$\sin^2 2\te_{\mu \tau}\simeq 1$, which means that the second mass eigenstate
is $\nu_2 \simeq \fr{1}{\sq{2}}(\nu_{\mu }+\nu_{\tau })$, and therefore the
$\nu_e$ state should oscillate with $50\%$ probability each into $\nu_{\mu}$
and $\nu_{\tau }$. However, at this level 
$\nu_e-\nu_2$ oscillation cannot occur because of zero mass squared
difference. Indeed, the neutrino mass spectra is

\beq
m_{\nu_1}=m_{\nu_2}=m~,~~~~~~~~m_{\nu_3}=0~,
\label{masses}
\eeq
(inverted hierarchy), and consequently:

\beq
\Delta m_{32}^2=m^2~,~~~~\Delta m_{21}^2=0~.
\la{splits}
\eeq
{}From (\ref{atmdat}) we have

\beq
m^2\sim m^2_{\rm atm}\sim 10^{-3}~{\rm eV}^2~.
\la{atmsc}
\eeq
In order to get a nonzero $\De m_{21}^2$, the texture (\ref{nu})
should be slightly modified:

\begin{equation}
\begin{array}{ccc}
&  {\begin{array}{ccc}
\hspace{-5mm}~~ & \,\, & \,\,

\end{array}}\\ \vspace{2mm}
\begin{array}{c}
\\  \\
\end{array}\!\!\!\!\!\hat{M}_{\nu }=&{\left(\begin{array}{ccc}
\,\,\rho_1~~ &\,\,m_1~~ &
\,\,m_2
\\
\,\,m_1~~   &\,\,\rho_2~~  &
\,\,\rho_4
 \\
\,\,m_2~~ &\,\,\rho_4~~ &\,\,\rho_3
\end{array}\right) }~,
\end{array}  \!\!~~~~~
\label{1nu}
\eeq
with small deviations $\rho_i \ll m_1, m_2$, in order to preserve bi-maximal
mixing. Of course, it is possible to choose $\rho_i$s by hand in such a
way
as to get the desirable picture. However, it would be much nicer to provide a natural mechanism for realizing this.  The goal is therefore to construct a model that will allow us to estimate values of $\rho_i$, which will give nonzero 
$\De m_{12}^2$, thereby guaranteeing $\nu_e-\nu_2$ oscillations. The value of
$\De m_{12}^2$ will select one of the currently viable oscillation scenarios for solar
neutrinos.
In \cite{barb}, the texture (\ref{nu}) was obtained by
imposing $L_e-L_{\mu }-L_{\tau }$ symmetry. Refs. \cite{josh} presented 
models in which non-zero  $\rho_i$ emerge at the two loop level and the 
large angle vacuum oscillation scenario
is realized.  In \cite{u1bimax} a ${\cal U}(1)$ flavor symmetry is used for
providing this bi-maximal mixing matrix.  In this paper we will follow 
the strategy of \cite{u1bimax}. Let us finally note that the texture in 
(\ref{nu}) is also discussed in \cite{bimax}.  
 
We introduce a ${\cal U}(1)$ flavor symmetry,
which distinguishes the families through the
prescription of ${\cal U}(1)$ charges. In many cases it turns out
that the ${\cal U}(1)$ is anomalous. It is well known that such anomalous
${\cal U}(1)$ factors can arise in string theories. Cancellation of the anomaly
occurs through the Green-Schwarz mechanism \cite{gs}. Due to the anomaly the
Fayet-Illiopoulos D-term
$\xi \int d^4\theta V_A$
is always generated, where in string theory $\xi $ is given by \cite{fi}

\begin{equation}
\xi =\frac{g_A^2M_P^2}{192\pi^2}{\rm Tr}Q~.
\label{xi}
\eeq
The $D_A$-term will have the form:

\begin{equation}
\frac{g_A^2}{8}D_A^2=\frac{g_A^2}{8}
\left(\Sigma Q_a|\varphi_a |^2+\xi \right)^2~,
\label{da}
\eeq
where $Q_a$ is the `anomalous' charge of $\varphi_a $ superfield.
For ${\cal U}(1)$ breaking we introduce the singlet superfield $X$ with 
${\cal U}(1)$ charge $Q_X$. Assuming
$\xi >0~$ [${\rm Tr}Q>0$ in (\ref{xi})],
and taking

\beq
Q_X=-1~,
\la{take}
\eeq
the cancellation of $D_A$ in (\ref{da}) and nonzero $\lan X\ran$ are
ensured:
$\langle X\rangle =\sqrt{\xi }$.
Further, we will take

\beq
\frac{\langle X\rangle }{M_P}\equiv \epsilon \simeq 0.2~,
\label{epsx}
\eeq
where $\ep $ turns out to be an important expansion parameter. Let us note that an anomalous ${\cal U}(1)$ for understanding the
hierarchies of fermion masses and mixings has been discussed in several papers 
of \cite{feru1}. In \cite{nuu1} a variety of neutrino
oscillation scenarios were constructed with the help of ${\cal U}(1)$.

Starting our investigation with the neutrino sector, let us first discuss
two ways of obtaining large/maximal neutrino mixings with the help
of ${\cal U}(1)$ flavor symmetry.
With two flavors of lepton doublets $l_1$ and $l_2$, one way of having large mixing is the so-called  {\it democratic approach}. 
In this case the ${\cal U}(1)$ symmetry does not distinguish the two flavors \cite{dem}, i.e. they have the same ${\cal U}(1)$ charges
$Q_{l_1}=Q_{l_2}=n$(positive integer number). In this case, the expected
neutrino mass matrix will be:

\beq
\begin{array}{cc}
&  {\begin{array}{cc}
\hspace{-5mm}~~ & \,\,

\end{array}}\\ \vspace{2mm}
\begin{array}{c}
\\  \\
\end{array}\!\!\!\!\!\hat{m}_{\nu }=&{\left(\begin{array}{cc}
\,\,1~~ &\,\,1
\\
\,\,1~~ &\,\,1
\end{array}\right)\bar m\ep^{2n} }~,~~~\bar m=\fr{h_u^2}{\bar M}~,
\end{array}  \!\!
\la{demnu}
\eeq
with entries of order unity ($\bar M$ is some mass scale and
we have assumed $Q_{h_u}=0$).
Therefore, naturally large
$\nu_1-\nu_2$ mixing is expected, $\sin^2 2\te_{12}\sim 1$. 
Also, one can expect $m_{\nu_1}\sim m_{\nu_2}$, and if this mechanism is
used for atmospheric neutrinos, somehow one has to keep one state light, 
in order to accomodate also the solar neutrino puzzle. This can be done
\cite{1N, su5u1} by introducing a single right handed neutrino 
${\cal N}$.
After integrating it out due to degeneracy, only one state acquires mass. The remaining
states can be used for the solar neutrino puzzle. An appropriate mass scale for
the latter can be generated by introducing a relatively heavy right handed state
${\cal N}'$ with suppressed coupling with ${\cal N}$.

A different approach is the so-called 
{\it maximal mixing mechanism} \cite{maxmix}. It is realized by
assigning different ${\cal U}(1)$ charges for the flavors 
$l_1, l_2$. Introducing two right handed states 
${\cal N}_1$, ${\cal N}_2$ and the following ${\cal U}(1)$ charge
prescriptions

$$
Q_{l_1}=k+n~,~~~Q_{l_2}=k~,~~~Q_{h_u}=0~,
$$
\beq
Q_{{\cal N}_1}=-Q_{{\cal N}_2}=k+k'~,
\la{maxQ}
\eeq
with $k, n, k'>0~, n\stackrel{>}{_{-}}k'$,
the `Dirac' and `Majorana' couplings will have the forms:

\beq
\begin{array}{cc}
 & {\begin{array}{cc}
{\cal N}_1~~~~&\,\,~~~{\cal N}_2 
\end{array}}\\ \vspace{2mm}
\begin{array}{c}
l_1\\ l_2 

\end{array}\!\!\!\!\! &{\left(\begin{array}{cc}
\,\, \epsilon^{2k+n+k'}~~ &
\,\, \epsilon^{n-k'}
\\
\,\, \epsilon^{2k+k'} ~~ &\,\,0
\end{array}\right)h_u }~,
\end{array}  \!\!~~~
\begin{array}{cc}
 & {\begin{array}{cc}
~{\cal N}_1~&\,\,
~~~{\cal N}_2~~~ 
\end{array}}\\ \vspace{2mm}
\begin{array}{c}
{\cal N}_1 \\ {\cal N}_2

\end{array}\!\!\!\!\! &{\left(\begin{array}{ccc}
\,\, \epsilon^{2(k+k')}
 &\,\,~~~1
\\
\,\, 1
&\,\,~~~0
\end{array}\right)M_{\cal N}~,
}
\end{array}~~~
\la{maxcoupl}
\eeq
After integrating out the heavy ${\cal N}_1, {\cal N}_2$ states, the
neutrino mass matrix is given by

\beq
\begin{array}{cc}
&  {\begin{array}{cc}
\hspace{-5mm}~~ & \,\,

\end{array}}\\ \vspace{2mm}
\begin{array}{c}
\\  \\
\end{array}\!\!\!\!\!\hat{m}_{\nu }=&{\left(\begin{array}{cc}
\,\,\ep^n~~ &\,\,1
\\
\,\,1~~ &\,\,0
\end{array}\right)\bar m }~,~~~\bar m=\fr{h_u^2\ep^{2k+n}}{M_{\cal N}}
\end{array}  \!\!~~~~~,
\la{maxnu}
\eeq
a quasi off-diagonal form, leading to a mixing angle

\beq
\sin^2 2\te_{12}=1-{\cal O}(\ep^{2n})~,
\la{maxmix}
\eeq
which is close to maximal mixing. The form (\ref{maxnu}) is
guaranteed
by the appropriate zero entries in (\ref{maxcoupl}), which are
ensured by ${\cal U}(1)$ symmetry. This mechanism turns out to be
very convenient for achieving nearly maximal mixings between neutrino
flavors within various realistic models, such as $SU(5)$
\cite{su5u1},
$SO(10)$ \cite{so10u1}, $SU(4)_c\tm SU(2)_L\tm SU(2)_R$ 
\cite{422}, etc.

Returning to our scheme, we attempt to obtain the bi-maximal 
texture (\ref{nu})
through ${\cal U}(1)$ flavor symmetry. For this, we will combine the two
mechanisms
for maximal mixing discussed above. Namely, the second and third lepton
doublet states will have the same ${\cal U}(1)$ charges, which will lead to their
large mixing. The state $l_1$ will have a suitable charge, one that ensures 
maximal $\nu_1 - \nu_2$ mixing.

Introducing two right handed ${\cal N}_{1, 2}$ neutrino states and
choosing ${\cal U}(1)$ charges as

$$
Q_X=-1~,~Q_{l_2}=Q_{l_3}=k~,~Q_{l_1}=k+n~,~Q_{h_u}=Q_{h_d}=0~,
$$
\beq
Q_{{\cal N}_1}=-Q_{{\cal N}_2}=k+k'~,
\label{charges}
\eeq
with

\beq
k, n, k'>0~,~~~~n\stackrel{>}{_{-}}k'~,
\la{condnk}
\eeq
the `Dirac' and `Majorana' couplings will have forms:

\begin{equation}
\begin{array}{cc}
 & {\begin{array}{cc}
{\cal N}_1~~&\,\,~~~{\cal N}_2~~
\end{array}}\\ \vspace{2mm}
\begin{array}{c}
l_1\\ l_2 \\ l_3

\end{array}\!\!\!\!\! &{\left(\begin{array}{ccc}
\,\, \epsilon^{2k+n+k'}~~ &
\,\, \epsilon^{n-k'}
\\
\,\, \epsilon^{2k+k'} ~~ &\,\,0
\\
\,\, \epsilon^{2k+k'} ~~ &\,\,0
\end{array}\right)h_u }~,
\end{array}  \!\!~~~
\begin{array}{cc}
 & {\begin{array}{cc}
{\cal N}_1~&\,\,
~~~{\cal N}_2~~~
\end{array}}\\ \vspace{2mm}
\begin{array}{c}
{\cal N}_1 \\ {\cal N}_2

\end{array}\!\!\!\!\! &{\left(\begin{array}{ccc}
\,\, \epsilon^{2(k+k')}
 &\,\,~~~1
\\
\,\, 1
&\,\,~~~0
\end{array}\right)M_{\cal N}~,
}
\end{array}~~~.
\label{Ns}
\eeq
After integrating out ${\cal N}_{1, 2}$, we obtain the texture

\begin{equation}
\begin{array}{ccc}
&  {\begin{array}{ccc}
\hspace{-5mm}~~ & \,\, & \,\,

\end{array}}\\ \vspace{2mm}
\begin{array}{c}
 \\  \\
 \end{array}\!\!\!\!\!\hat{M}_{\nu }\propto &{\left(\begin{array}{ccc}
\,\,\epsilon^n~~ &\,\,1~~ &
\,\,1
\\
\,\,1~~   &\,\,0~~  &
\,\,0
 \\
\,\,1~~ &\,\,0~~ &\,\,0
\end{array}\right) }m~,~~~~~~m=\frac{\epsilon^{2k+n}h_u^2}{M_{\cal N}}~,
\end{array}  \!\!
\label{u1nu}
\eeq
which differs from (\ref{nu}) by a non-zero (1, 1) element.
In (\ref{u1nu}) coefficients of order
unity are assumed. The nonzero (1, 1) entry in (\ref{u1nu}) 
guarantees that $\De m_{12}^2 {\neq}0$. 
Using (\ref{defamp}) and (\ref{u1nu}) the oscillation parameters are:

$$
\Delta m^2_{32}\equiv m_{\rm atm}^2= m^2\sim 10^{-3}~{\rm eV}^2~,
$$
\beq
{\cal A}(\nu_{\mu }\to \nu_{\tau })\sim 1~,
\label{atmosc}
\eeq

$$
\Delta m^2_{21 }\simeq 2m_{\rm atm}^2\epsilon^n~,
$$
\beq
{\cal A}(\nu_e \to \nu_{\mu , \tau }) =1-{\cal O}(\epsilon^{2n})~.
\label{solosc}
\eeq
Note, that the model does not constrain $n$ for the time being. So, 
LAMSW, LOW MSW and LAVO solutions for solar neutrinos, can be realized.
With prescription (\ref{charges}), the expected contribution from the
charged
lepton sector to the angles $\te_{23}^l$ and $\te_{12}^l$ will be
$\sim 1$ and $\sim \ep^n$ respectively. These do not change 
the form of $V_{\nu }$ in (\ref{lepckm}).

The ${\cal U}(1)$ charge selection in (\ref{charges}) nicely blends with
the charged
fermion sector. Indeed, considering the following prescription:

$$
Q_{q_3}=0~,~~Q_{q_2}=2~,~~Q_{q_1}=3~,~~
Q_{d^c_3}=Q_{d^c_2}=p+k~,~~
$$
$$
Q_{d^c_1}=p+k+2~,~~Q_{u^c_3}=0~,~~Q_{u^c_2}=1~,~~Q_{u^c_1}=3~,
$$
\beq
Q_{e_3^c}=p~,~~~Q_{e_2^c}=p+2~,~~~
Q_{e_1^c}=p+5-n~,~~~
\label{qch}
\eeq
the structures of Yukawa matrices, for up-down quarks and charged leptons
 respectively:

\begin{equation}
\begin{array}{ccc}
 & {\begin{array}{ccc}
\hspace{-5mm} u^c_1 & \,\,~u^c_2 ~~ & \,\,u^c_3 ~

\end{array}}\\ \vspace{2mm}
\begin{array}{c}
q_1 \\ q_2 \\q_3
 \end{array}\!\!\!\!\! &{\left(\begin{array}{ccc}
\,\,\epsilon^6~~ &\,\,\epsilon^4~~ &
\,\,\epsilon^3  
\\  
\,\,\epsilon^5~~   &\,\,\epsilon^3~~  &
\,\,\ep^2
 \\
\,\,\epsilon^3~~ &\,\,\epsilon ~~ &\,\,1
\end{array}\right)h_u }~, 
\end{array}  \!\!  ~~~~~
\label{up}
\eeq

\begin{equation}
\begin{array}{ccc}
 & {\begin{array}{ccc}
\hspace{-5mm} d^c_1~ & \,\,d^c_2 ~~ & \,\,d^c_3 ~~~~~~

\end{array}}\\ \vspace{2mm}
\begin{array}{c}  
q_1 \\ q_2 \\q_3
 \end{array}\!\!\!\!\! &{\left(\begin{array}{ccc} 
\,\,\epsilon^5~~ &\,\,\epsilon^3~~ &
\,\,\epsilon^3
\\
\,\,\epsilon^4~~   &\,\,\epsilon^2~~  &
\,\,\epsilon^2
 \\
\,\,\epsilon^2~~ &\,\,1~~ &\,\,1
\end{array}\right)\epsilon^{p+k}h_d }~,
\end{array}  \!\!  ~~~~~
\label{down}
\eeq

\begin{equation}
\begin{array}{ccc}
 & {\begin{array}{ccc}
\hspace{-7mm} e^c_1~~~~~ & \,\,e^c_2 ~~~ & \,\,e^c_3 ~~
  
\end{array}}\\ \vspace{2mm}
\begin{array}{c}
l_1 \\ l_2 \\l_3
 \end{array}\!\!\!\!\! &{\left(\begin{array}{ccc}
\,\,\epsilon^5~~ &\,\,\epsilon^{n+2}~~ &
\,\,\epsilon^n
\\  
\,\,\epsilon^{5-n}~~   &\,\,\epsilon^2~~  &
\,\,1
 \\
\,\,\epsilon^{5-n}~~ &\,\,\epsilon^2~~ &\,\,1
\end{array}\right)\epsilon^{p+k}h_d }~.
\end{array}  \!\!  ~~~~~
\label{lept}
\eeq
Upon diagonalization of (\ref{up})-(\ref{lept}) it is easy to verify that
the desired relations (\ref{ulam})-(\ref{ckm}) for the Yukawa couplings and CKM
matrix elements are realized. From (\ref{down}), (\ref{lept}) we have

\beq
\tan \bt \sim \ep^{p+k}\fr{m_t}{m_b}~.
\label{tangens}
\eeq
As we previously mentioned, MSSM does not fix the values of $n, k, p$ in
(\ref{charges}), (\ref{qch}). Because of this, the solar neutrino oscillation
scenario is not
specified. According to (\ref{solosc}) there is possibility for all three:
LAMSW, LOW MSW and LAVO solutions. Namely, for
$n=3$ we have $\De m_{12}^2\sim 10^{-5}~{\rm eV}^2$, which corresponds to
LAMSW. $n=6$ give $\De m_{12}^2\sim 10^{-7}~{\rm eV}^2$, which is the scale
for LOW MSW, while $n=10$ generates the scale
$\De m_{12}^2\sim 10^{-10}~{\rm eV}^2$, corresponding to LAVO
solution.  It would be interesting to look for models/scenarios, which somehow fix
the values of $n, p, k$, that dictate the solar neutrino oscillation scenario, Here
we present two cases in which this happens.  

\vspace{0.5cm}

Consider SUSY $SU(5)$ GUT. Its matter sector consists of $10+\bar 5$
supermultiplets per generation. Due to these unified multiplets:

\beq
Q_q=Q_{e^c}=Q_{u^c}=Q_{10}~,~~~~~
Q_l=Q_{d^c}=Q_{\bar 5}~.
\label{chsu5}
\eeq
Hierarchies of the CKM matrix elements in (\ref{ckm}) dictate the
relative ${\cal U}(1)$ charges of the $10$-plets

\beq
Q_{10_3}=0~,~~~Q_{10_2}=2~,~~~Q_{10_1}=3~,
\label{ch10}
\eeq
while the Yukawa hierarchies (\ref{ulam})-(\ref{elam}), together with
(\ref{ch10}), require that

\beq
Q_{\bar 5_3}=Q_{\bar 5_2}=k~,~~~~Q_{\bar 5_1}=k+2~.
\label{ch5}
\eeq
Comparing (\ref{chsu5})-(\ref{ch5}) with (\ref{charges}),
(\ref{qch}) we see that the minimal $SU(5)$ GUT fixes $n$ and $p$
as

\beq
n=2~,~~~~~p=0~.
\label{nk}
\eeq
The mass squared splitting in 
(\ref{solosc}) then equals 
$\De m_{12}^2\sim 10^{-4} {\rm eV}^2$, which is a reasonable scale for 
LAMSW scenario. 
Therefore, realisation of our bi-maximal mixing scenario in
the framework of $SU(5)$ GUT dictates that the LAMSW scenario is responsible for
the solar neutrino deficit. The same conclusion can be reached for 
$SO(10)$ GUT where we have three $16$-plets of the chiral 
supermultiplets which unify the quark-lepton superfields. We do not present the
details here but refer the reader to \cite{so10u1}, where an
explicit $SO(10)$ model with anomalous ${\cal U}(1)$
flavor symmetry is considered  for explanations of fermion masses, their mixings,
as well as neutrino anomalies.

For our second example,  
we present a model in which anomalous flavor 
${\cal U}(1)$ also mediates SUSY breaking \cite{susyu1}. In this 
case soft masses ($\sim m_S$), emerging through non-zero $D_A$-term,
dominate over $m_{3/2}$ (gravitino mass). This fact can be used for
natural suppression of FCNC through the so-called decoupling solution
\cite{FCNC}. The generated soft mass squared for the scalar
component of superfield $\phi_a$ is

\beq
m^2_{\tilde{\phi}_i}=m_S^2Q_a~.
\label{soft}
\eeq
Therefore, sparticles
with non-zero ${\cal U}(1)$ charges will be relatively heavy.  
A scale $\sim 10$~TeV is enough for adequate
suppression of all FCNC processes \cite{mas}.
It turns out that in this scenario there also occurs suppression of
dimension five nucleon decay \cite{d5nel, FCd5}.
In \cite{FCd5} we presented a specific model of anomalous flavor
${\cal U}(1)$ mediated SUSY breaking. 
Suppressions of FCNC and $d=5$
nucleon decay were guaranteed through 
heavy ($\sim 10$~TeV) first and second sparticle generations
(and also through heavy $\tilde{b}$, $\tilde{\tau }$, which require the
low $\tan \bt $ regime). 

In order to have possitive soft squared masses for squarks and sleptons,
they should have charges $Q_{\tl{q}, \tl{l}}>0$ [see (\ref{soft})]. On the
other hand, 
from (\ref{solosc}), the realization of LAVO requires $n=10$, which for
$Q_{e^c}$ gives a negative ${\cal U}(1)$ charge [see (\ref{qch}).
$p+k\stackrel{<}{_{-}}3$, since 
$10^{-2}\stackrel{<}{_{\sim }}\lam_{b, \tau }\stackrel{<}{_{\sim }}1$]
and therefore negative $m_{\tl{e^c}}^2$. However, the value $n=3$, which is needed
for LAMSW, is possible. Also, the case $n=6$ is possible if 
$p\stackrel{>}{_{-}}1$, in order to guarantee $Q_{e^c}\stackrel{>}{_{-}}0$
in (\ref{qch}). This gives the LOW MSW solution. Note also, that
in this particular case the value of $\tan \bt $ 
in (\ref{tangens}) is either intermediate or low. 
Therefore, the scenario in which anomalous
flavor
${\cal U}(1)$ mediates SUSY breaking permits  LAMSW and LOW MSW  
oscillations for solar neutrinos, but excludes LAVO even in the framework
of MSSM.

\vspace{0.3cm}

As we have seen the texture (\ref{u1nu}) provides bi-maximal neutrino
mixing. 
However, $\hat{M}_{\nu }$ has the form (\ref{u1nu}) if we neglect
renormalizations. In order for the analysis to be  complete,  we
should take this into account.
Crucial for this are the structures in (\ref{Ns}), which
are prescribed at scale $M_X$ (${\cal U}(1)$ summetry breaking
scale). In the ranges $M_X - M_{{\cal N}}$ and $M_{{\cal N}} - M_Z$,
renormalization effects will occur and one has to make sure that
the successful picture of bi-maximal mixing will not be spoiled.
Let us now confirm that this is indeed the case here.

Between $M_X$ and $M_{{\cal N}}$ the states ${\cal N}_{1, 2}$ are not
decoupled and we have to renormalize the following couplings:

\beq
l\hat{\lam }_{\nu }{\cal N}h_u+{\cal N}\hat{M}{\cal N}~.
\la{nucoupl}
\eeq
Renormalization group equations (RGE) for the elements of $\hat{\lam
}_{\nu }$ and  $\hat{M}$ respectively are:

\beq
16\pi^2\fr{d\lam^{ij}_{\nu }}{dt}=
\l -c_ag_a^2\lam_{\nu }+2\lam_{\nu }\lam_{\nu }^T\lam_{\nu }+
\lam_e\lam_e^T\lam_{\nu }+
3{\rm tr}(\lam_u \lam_u^T)\lam_{\nu }\r^{ij}~,
\la{nurge}
\eeq

\beq
16\pi^2\fr{d}{dt}M^{ij}=
\l M\lam_{\nu }^T\lam_{\nu }+\lam_{\nu }^T\lam_{\nu }M\r^{ij}~,
\la{Nrge}
\eeq
where $t=\ln \mu$ is a renormalization scale factor and $g_a$ ($a=1, 2, 3$)
are the gauge couplings of $U(1)_Y, SU(2)_L$ and $SU(3)_c$ respectively.
For MSSM $c_a=(\fr{3}{5},~3,~0)$.

We will work in a basis in which $\hat{\lam }_e$ is a diagonal. 
The lepton CKM matrix is then completely determined
through the neutrino mixing matrix.
In this basis, instead of (\ref{Ns}), 
$\hat{\lam }_{\nu }$ has the form

\begin{equation}
\begin{array}{cc}
\hat{\lam }_{\nu }=\hs{-0.3cm} &{\left(\begin{array}{ccc}
\,\, \epsilon^{2k+n+k'}~~ &
\,\, \epsilon^{n-k'}
\\
\,\, \epsilon^{2k+k'} ~~ &\,\,\ep^{2n-k'}
\\
\,\, \epsilon^{2k+k'} ~~ &\,\,\ep^{2n-k'}
\end{array}\right) }~.
\end{array}  \!\!~~~
\label{nulam}
\eeq
It is easy to see that the non-zero but suppressed (2, 2) and (3, 2) entries 
in (\ref{nulam})  
do not change results. The important thing is that, after 
renormalization,

\beq
\de_{2 (3)}\equiv \fr{\lam^{22(32)}_{\nu }}{\lam^{12}_{\nu }}
\stackrel{<}{_{\sim }}\ep^{n}~.
\la{condel}
\eeq
Also, in the texture for ${\cal N}_{1, 2}$ in (\ref{Ns}), the elements
$M^{11}$, $M^{22}$ must satisfy 

\beq
\De_{1 (2)} \equiv \fr{M^{11 (22)}}{M^{12}}\ll 1~.
\la{condDe}
\eeq
At scale $M_X$ the conditions (\ref{condel}), (\ref{condDe}) are
guaranteed by ${\cal U}(1)$,
and we have to make sure that $\de_2, \de_3, \De_1, \De_2$
do not get significant contributions from renormalizations.
{}From (\ref{nurge}), (\ref{Nrge}) the RGEs for $\de $ and $\De $, 
to a good approximation, are:

\beq
16\pi^2\fr{d\de_2}{dt}\simeq (\lam_{\mu }^2+4\ep^{4k+2k'})\hs{1mm}\de_2+
2\ep^{4k+2k'+n}~,
\la{de2rge}
\eeq

\beq
16\pi^2\fr{d\de_3}{dt}\simeq (\lam_{\tau }^2+4\ep^{4k+2k'})\hs{1mm}\de_3+
2\ep^{4k+2k'+n}~,
\la{de3rge}
\eeq

\beq
16\pi^2\fr{d\De_{1}}{dt}\simeq (2\ep^{4k+2k'}-\ep^{2n-2k'})\hs{1mm}\De_1+
6\ep^{2k+2n}~,
\la{De1rge1}
\eeq

\beq
16\pi^2\fr{d\De_{2}}{dt}\simeq (\ep^{2n-2k'}-2\ep^{4k+2k'})\hs{1mm}\De_2+
6\ep^{2k+2n}~.
\la{De2rge1}
\eeq
The approximate solutions of (\ref{de2rge})-(\ref{De2rge1}) are

\beq
\de_2\simeq \fr{1}{16\pi^2}[(\lam_{\mu }^2+4\ep^{4k+2k'})\hs{1mm}\de_2^0+
2\ep^{4k+2k'+n}]
\ln \fr{M_{\cal N}}{M_X}+\de_2^0\stackrel{<}{_{\sim }}
\ep^n~,
\la{solde2}
\eeq

\beq
\de_3\simeq \fr{1}{16\pi^2}[(\lam_{\tau }^2+4\ep^{4k+2k'})\hs{1mm}\de_3^0+
2\ep^{4k+2k'+n}]
\ln \fr{M_{\cal N}}{M_X}+\de_3^0\stackrel{<}{_{\sim }}
\ep^n~,
\la{solde3}
\eeq

\beq
\De_1\simeq
\De_1^0+\fr{1}{16\pi^2}[(2\ep^{4k+2k'}-\ep^{2n-2k'})\hs{1mm}\De_1^0+
6\ep^{2k+2n}]\ln \fr{M_{\cal N}}{M_X}~,
\la{solDe1}
\eeq

\beq
\De_2\simeq \fr{3}{8\pi^2 }\ep^{2k+2n}\ln \fr{M_{\cal N}}{M_X}~,
\la{solDe2}
\eeq
where $\de^0$, $\De^0$ denote their values at $M_X$, while
$\de $, $\De $ in (\ref{solde2})-(\ref{solDe2}) are their values
at $M_{\cal N}$ (at $M_X$, $\De_2^0=0$ due
to ${\cal U}(1)$).
Equations (\ref{solde2})-(\ref{solDe2}) convince us that down to scale $M_{\cal
N}$, the conditions (\ref{condel}), (\ref{condDe}) are easily satisfied 
even for $M_X/M_{\cal N}\sim 10^{15}$.

At scale $M_{\cal N}$ the states ${\cal N}_{1, 2}$ decouple and
$\hat{M}_{\nu }$ is generated. Therefore, below $M_{\cal N}$ , $\hat{M}_{\nu }$ will run through appropriate $d=5$
operators. The RGE for $\hat{M}_{\nu }$ is:

\beq
16\pi^2\fr{d}{dt}\hat{M}_{\nu}=-\bar c_ag_a^2\hat{M}_{\nu}+
\hat{\lam }_e\hat{\lam }_e^T\hat{M}_{\nu}+
\hat{M}_{\nu}\hat{\lam }_e\hat{\lam }_e^T+
6{\rm tr}(\lam_u^T\lam_u)\hat{M}_{\nu}~,
\la{d5nurge}
\eeq
where $\bar c_a=(\fr{6}{5},~6,~0)$.

The important point is that in (\ref{u1nu}) the elements (1, 1) (2, 3),
(3, 3),
(2, 2)$\stackrel{<}{_{\sim }}m\ep^{n}$.  Using the notations

\beq
\fr{M_{\nu }^{11}}{M_{\nu }^{12}}=X_1~,~~
\fr{M_{\nu }^{22}}{M_{\nu }^{12}}=X_2~,~~
\fr{M_{\nu }^{33}}{M_{\nu }^{12}}=X_3~,~~
\fr{M_{\nu }^{23}}{M_{\nu }^{12}}=X_4~,
\la{defX}
\eeq
the conditions 

\beq
X_i\stackrel{<}{_{\sim }}\ep^n
\la{condX}
\eeq
must be satisfied at scale $M_Z$. If in 
(\ref{defX}) $M_{\nu}^{12}$ will be replaced by $M_{\nu}^{13}$, the
conditions (\ref{condX}) will still occur, since we require that
$M_{\nu}^{12}\sim M_{\nu}^{13}$. Let us first demonstrate that
the ratio $M_{\nu}^{12}/M_{\nu}^{13}\equiv r$ does not change significantly
under renormalization. Using (\ref{d5nurge}), the RGE for $r$:

\beq
16\pi^2\fr{d\ln r}{dt}\simeq -\lam_{\tau }^2~,
\la{rrge}
\eeq  
with the approximate solution

\beq
r\simeq r^0(1-\fr{\lam_{\tau }^2}{16\pi^2}\ln \fr{M_Z}{M_{\cal N}})
\sim r^0~,
\la{solr}
\eeq
where $r$ and $r^0$ are the values at scales $M_{\cal N}$ and $M_Z$
respectively.
(\ref{solr}) demonstrates that the magnitude of $r$ is not significantly altered.
The RGEs for $X_i$, to good approximations, are:

$$
16\pi^2\fr{d\ln X_1}{dt}\simeq -\lam_{\mu }^2~,~~~~~
16\pi^2\fr{d\ln X_2}{dt}\simeq \lam_{\mu }^2~,
$$
\beq
16\pi^2\fr{d\ln X_3}{dt}\simeq 2\lam_{\tau }^2~,~~~~~
16\pi^2\fr{d\ln X_4}{dt}\simeq \lam_{\tau }^2~,
\la{X3rge}
\eeq
with approximate solutions

$$
X_{1 (2)}\simeq X_{1 (2)}^0\l 1\mp \fr{\lam_{\mu }^2}{16\pi^2}
\ln \fr{M_Z}{M_{\cal N}}\r \simeq X_{1 (2)}^0~,
$$
\beq
X_3\simeq X_3^0\l 1+\fr{\lam_{\tau }^2}{8\pi^2}
\ln \fr{M_Z}{M_{\cal N}}\r \sim X_3^0~,~~~
X_4\simeq X_4^0\l 1+\fr{\lam_{\tau }^2}{16\pi^2}
\ln \fr{M_Z}{M_{\cal N}}\r \sim X_4^0~,
\la{solX3}
\eeq
which demonstrates that the conditions in (\ref{condX}) are satisfied
since $X_i$ is proportional to $X_i^0$, which does not exceed
$\ep^n$.

We therefore can conclude that renormalization effects do not significantly affect
$\hat{M}_{\nu }$ which has the desired  
form (\ref{u1nu}) at scale $M_Z$. Let us note that the papers in \cite{efren}
investigate the influence of renormalizations
on neutrino mixings and oscillations, while 
\cite{stab} present models with large neutrino mixings, that are stable against
radiative corrections.

\vspace{0.3cm}

In conclusion, within MSSM and beyond we have addressed the problem of flavor
and
neutrino anomalies. For a simultaneous resolution an anomalous 
${\cal U}(1)$ flavor symmetry was invoked. Bi-maximal neutrino mixing
texture was generated and the observed hierarchies between charged fermion
masses and their mixings were obtained.
Renormalization group analysis shows that the
bi-maximal neutrino mixing picture is stable against quantum 
corrections.

\vspace{0.5mm}
Z.T. would like to thank the Organizers of NATO 2000 meeting
for warm hospitality at Cascais - Portugal and for their support.

\bibliographystyle{unsrt}

\end{document}